\documentclass[prc,preprint]{revtex4}
\usepackage{amsmath,amssymb,epsfig,bm}
\newcommand{\be}{\begin{equation}}
\newcommand{\ee}{\end{equation}}
\newcommand{\bea}{\begin{eqnarray}}
\newcommand{\eea}{\end{eqnarray}}

\newcommand{\anu}{\bar\nu}

\newcommand{\sla}{\! \not \!}
\newcommand{\ep}{\epsilon}
\newcommand{\Real}{{\rm Re}}
\newcommand{\Img}{\Im{\rm m}}
\newcommand{\vecq}{\bm q}
\newcommand{\vecp}{\bm p}
\newcommand{\veck}{\bm k}
\newcommand{\dF}{F^{\dagger}}
\newcommand{\vs}{\varsigma}
\newcommand{\sigmavec}{{\bm \sigma}}

\begin{document}
\title{
Vertex renormalization in weak decays of Cooper pairs 
and cooling  compact stars
}
\author{Armen Sedrakian$^{1,2}$, Herbert M\"uther$^1$, and Peter Schuck$^3$
}
\address{$^1$Institute for Theoretical Physics, \\
T\"ubingen University, D-72076 T\"ubingen, Germany\\
$^2$Institute for Theoretical Physics, \\
J. W. Goethe University, D-60438  Frankfurt-Main, Germany\\
$^3$Groupe de Physique Th\'eorique, Institut de Physique Nucl\'eaire,
            91406 Orsay, France
}
\date{\today}
 
\begin{abstract}                           
At temperatures below the critical temperature of superfluid 
phase transition baryonic matter emits neutrinos by breaking 
and recombination of Cooper pairs formed in the condensate. 
The strong interactions in the nuclear medium modify the weak 
interaction vertices and the associated neutrino loss rates. 
We study these modifications non-perturbatively by summing 
infinite series of particle-hole loops in the $S$-wave 
superfluid neutron matter. The pairing and particle-hole 
interactions in neutron matter are described in the framework 
of the BCS and Fermi-liquid theories derived from microscopic 
interactions. Consistent with the $f$-sum rule, 
the leading order contribution to the polarization tensor 
arises at $O(q^2)$ in the small momentum transfer, $q$, expansion. 
The associated neutrino emission rate is parametrically suppressed 
compared to its one-loop counterpart by a factor of the order 
of $5\times 10^{-3}$, the parameter being the baryon recoil 
in units of temperature.

\end{abstract}

\maketitle

\section{Introduction}
\label{sec:1}

Pair-correlated baryonic matter in compact stars emits neutrinos via the 
weak neutral current processes of pair-breaking and 
recombination~\cite{FRS78,VS87+MIGDAL90}
\be\label{eq:1} 
\{NN\} \to N+N+\nu_f+\anu_f, \quad N+N\to \{NN\} + \nu_f + \anu_f,
\ee
where $\{NN\}$ refers to a Cooper pair, $N+N$ to two quasiparticle
excitations, $\nu_f$ and $\anu_f$ to the neutrino and anti-neutrino 
of flavor $f$. The process (\ref{eq:1}) is limited to 
the temperature domain   $T^*\le T\le T_c$,  where 
$T_c$ is the critical temperature of pairing phase transition and 
$T^*\sim 0.2~T_c $. At and above $T_c$ this reaction cannot occur,
since momentum and energy cannot be conserved simultaneously
in a process $N \to N+ \nu_f + \anu_f$, i.~e., an on-mass-shell 
fermion cannot produce bremsstrahlung (in the absence of external 
gauge fields). At asymptotically low temperatures, $T\le T^*$, 
the rate of the process (\ref{eq:1}) is exponentially small 
since the number of excitations out of the condensate is suppressed  
as ${\rm exp}(-\Delta/T)$, where $\Delta(T)$ is the gap 
in the quasiparticle spectrum.

Cooling simulations of neutron stars revealed the efficiency 
of the processes (\ref{eq:1})  in refrigerating the baryonic 
interiors of a compact star from temperatures $T\le T_c\sim 10^9$~K
down to temperatures of the order of $10^8$ 
K~\cite{SCHAAB96,TSU,GRIG,PAGE_MINIMAL}.
The temperature domain above corresponds to 
the {\it neutrino cooling era}  
that spans the time domain $10^2\le t\le 10^5$ years. The predicted 
surface temperatures of neutron stars during this and the following 
{\it photon cooling era}  (where the star loses its thermal 
energy by emission of photons from the surface) are sensitive 
to the neutrino emission rates within this time-domain.
Remarkably, the process (\ref{eq:1})
relates the cooling rate of a compact star to the microscopic physics
of its interiors and is particularly sensitive to the density-temperature 
phase diagram of paired baryonic matter. 
Therefore, the measurements of the surface  (photon) 
luminosities of neutron stars and their interpretation 
in terms of cooling simulations have predictive power for analyzing 
the phase diagram and composition of baryonic 
matter~\cite{GRIG_VOSK,KHODEL,VOSK_REVIEW,SEDRAKIAN_REVIEW}.

The rate of the process (\ref{eq:1}) was computed independently
(and within alternative methods)  in Refs.~\cite{FRS78,VS87+MIGDAL90}  
in the case where the pairing interaction is in the $^1S_0$ partial wave, 
i.~e., nucleons are paired in a  spin-0, isospin-1 state, (the influence
of electric charge carried by proton Cooper pairs and the case of pairs 
forming a spin 1-superfluid are discussed in Refs.~\cite{1LOOP}).
In propagator language these rates correspond to the {\it one-loop}  
approximation to the polarization tensor of baryonic 
matter~\cite{VS87+MIGDAL90,SEDRAKIAN_REVIEW}. It has been known for a long
time that the gauge invariance requires summation of infinite series 
of loop diagrams (known as RPA resummation) 
in conventional superconductors~\cite{NAMBU,SCHRIEFFER}. 
Ref.~\cite{KUNDU_REDDY} carried out RPA  summations in neutron
and quark matter within the context and kinematics of neutrino 
scattering in hot proto-neutron stars. Subsequently, Ref.~\cite{LP} 
applied the gauge invariance and vertices derived by 
Nambu~\cite{NAMBU} to the neutrino emission processes in the 
vector channel and concluded that the rates are suppressed by 
by many order of magnitude.

We shall arrive below at results which are at variance 
to those of  Ref.~\cite{LP} for the following reasons.
The effective vertices implemented in Ref.~\cite{LP} were derived within a 
zero-temperature theory~\cite{NAMBU}; herein we shall derive the 
vertices at finite temperature, consistent with the polarization 
tensor of matter. This guarantees that the unitarity of the $S$-matrix
in the quantum mechanical process of the bremsstrahlung is preserved.
Second, while in Ref.~\cite{LP} the matrix elements
are expanded in the parameter $v_F/c$, where $v_F$ is the baryon velocity,
and the leading order contribution to the rates comes from the 
terms $O(v_F/c)^4$, herein the polarization tensor is expanded 
in small momentum transfer, $q$; 
the leading order contribution to the rates arises
at $O(q^2)$ consistent with the $f$-sum rule. 
As a consequence, we find a suppression of the neutrino bremsstrahlung
rate which is by several orders of magnitude less than predicted in 
Ref.~\cite{LP}. Since the magnitude of the pairing gaps are not well 
known and the neutrino bremsstrahlung rates are rather sensitive to 
the value of the gap, the pairing braking process in $S$-wave superfluids
remains a potentially viable mechanism of cooling of neutron stars.

We start by setting up a formalism for computing the vertex corrections
to weak reaction rates that arise due to the strong interactions 
in baryonic matter. We consider the case of neutron matter 
at subnuclear densities, which we describe within the Landau 
Fermi-liquid theory derived from microscopic interactions. 
At these  densities neutron matter is characterized by an 
isotropic order parameter arising form the interaction in 
the $^1S_0$ partial wave channel; we solve the corresponding 
problem of pairing in the framework of the Bardeen-Cooper-Schrieffer 
(BCS) theory. In the non-relativistic limit, the vector and axial 
vector weak vertices are associated with scalar and spinor 
perturbations. Thus, the problem of the weak vertex renormalization 
reduces to a study of the effective three-point vertices that sum-up 
particle-hole irreduceable ladders in superfluid matter in the scalar 
and spin channels. In a wider context, such resummations 
describe of the low-frequency, 
long-wave-length collective modes of superfluid
Fermi-liquids~\cite{SCHRIEFFER,LEGGET,MIGDAL}.  In particlar, 
such resummations, 
known as quasiparticle random phase approximation (QRPA), 
have been widely used in nuclear structure calculations~\cite{RING_SCHUCK}.

The paper is organized as follows. Section~\ref{sec:2} introduces the 
Green's functions formalism and computes, for the sake of illustration, 
the neutrino rates at one-loop. In Sec.~\ref{sec:3} the modifications 
of the weak interaction vertices are computed by summing  
particle-hole ladders in superfluid neutron matter.
Section~\ref{sec:4} is devoted to the computation of the full polarization 
tensor that includes the vertex corrections derived in the preceding 
Section. Section~\ref{sec:5} contains our conclusion. Some technical 
details are relegated to the Appendix.

\section{$S$-wave pair-condensation in neutron matter}
\label{sec:2}

Below we shall describe 
the neutron pair-condensate at subnuclear densities within
the framework of the Fermi-liquid theory, which assumes that the 
elementary degrees of freedom are quasiparticles with well defined 
momentum-energy relation and infinite life-time. The interactions between
the quasiparticles are then described in terms of Landau parameters
which depend on the momentum transfer (or scattering angle). Since the 
scattering angles involved are typically small, the momentum dependence 
of Landau parameters is further approximated by the leading and 
next-to-leading order terms in the expansion in Legendre polynomial with
respect to the scattering angle. The problem of pairing in neutron matter 
will be solved below within the BCS approximation, where the anomalous 
self-energy (the gap function) is computed from the bare 
interaction while the normal self-energy is computed within the decoupling
approximation which ignores the effects of pair-correlations on the 
single particle spectrum of quasiparticles. A number of factors such 
as the renormalization of the pairing 
interaction~\cite{PETHICK,CLARK76,CHEN86,CHEN93,WAMBACH,SCHULZE96,SEDRAKIAN03,SCHWENK,LOMB1,FABROCINI}
and the wave-function 
renormalization~\cite{SEDRAKIAN03,LOMB2,MUDICK}, 
 affect the absolute value of the gap. 
The role of these factors has not been settled yet, and we shall 
employ  below the standard BCS approach which has led to convergent 
and verifiable results for neutron matter (see the 
reviews~\cite{REVIEW1,REVIEW2,REVIEW3}).

Since the baryonic component of stellar matter is in thermal 
equilibrium to a good approximation, we shall adopt the Matsubara Green's 
functions for the description of the neutron condensate 
and for the evaluation of the polarization tensor. In the case of $^1S_0$ 
pairing these are defined as (e.~g. Ref.~\cite{ABRIKOSOV}, pg. 120)
\bea
\label{eq:P1}
\hat G_{\alpha\alpha'}(\vecp ,\tau) &=& -\delta_{\alpha\alpha'}
\langle T_{\tau} a_{p\alpha}(\tau)a^{\dagger}_{p\alpha'}(0)\rangle ,
 \\
\label{eq:P2}
\hat F_{\alpha\alpha'}(\vecp ,\tau) &=& \delta_{\vs\vs '}
\langle T_{\tau} a_{-p\downarrow}(\tau)
a_{p\uparrow}(0)\rangle , \\
\label{eq:P3}
\hat F^{\dagger}_{\alpha\alpha'}(\vecp ,\tau) &=& \delta_{\vs\vs '}
\langle T_{\tau} a_{p\uparrow}^{\dagger}(\tau)
a^{\dagger}_{-p\downarrow}(0)\rangle ,
\eea
where $\alpha$ stands for spin $\sigma = \uparrow , \downarrow$ and isospin 
$\vs$,  $\tau$ is the imaginary time, $T_{\tau}$ is the imaginary time
ordering symbol, $a^{\dagger}_{p\sigma}(\tau)$ and $a_{p\sigma}(\tau)$ 
are the creation and destruction operators.
In the momentum representation the propagators are given by 
\bea \label{P5}
\hat G_{\alpha\alpha'}(ip_n,\vecp) 
&=& \delta_{\sigma\sigma'}\delta{\vs\vs'}\left(
\frac{u_p^2}{ip_n-\ep_p} +\frac{v_p^2}{ip_n+\ep_p} \right),\\
     \label{P6} 
\hat F_{\alpha\alpha'}(ip_n,\vecp) 
&=& - i\sigma_y\delta{\vs\vs'}
              u_pv_p\left(\frac{1}{ip_n-\ep_p}-\frac{1}{ip_n+\ep_p}\right),
\eea
and $ F_{\alpha\alpha'}^{\dagger}(ip_n,\vecp) 
= F_{\alpha\alpha'}(ip_n,\vecp)$,  where $p_n = (2n+1)\pi T$ 
is the fermionic Matsubara frequency, $\sigma_y$ is the $y$-component
of the Pauli-matrix, $u_p^2 = (1/2)(1+\xi_p/\ep_p)$ 
and $v_p^2 = 1-u_p^2$ are the Bogolyubov amplitudes and 
\be\label{SPEC}
\ep_p= \sqrt{\xi_p^2+\Delta_p^2}
\ee
is the quasiparticle spectrum, where $\xi_p = p^2/2m +{\rm Re}\Sigma(p) -\mu$
is the spectrum in the unpaired state with $m$ and $\mu$ being the 
bare mass and chemical potential. Here $\Sigma(p)$ and 
$\Delta(p)$ are the normal and anomalous self-energies. 
For the later diagrammatic analysis we shall need the hole 
propagator which is given by 
\bea
\hat G_{\alpha\alpha'}^{\dagger}(ip_n,\vecp) 
&=&
\hat G_{\alpha\alpha'}(-ip_n,-\vecp) \nonumber\\
&=&
-\delta_{\sigma\sigma'}\delta{\vs\vs'}\left(
\frac{u_p^2}{ip_n+\ep_p} +\frac{v_p^2}{ip_n-\ep_p}\right) .
\eea
The  spin and isospin dependence of propagators for $S$-wave 
spin-0 and isospin-1 pairing is:
$\hat G_{\alpha\alpha'}(ip_n,\vecp)=\delta_{\sigma\sigma'}\delta{\vs\vs'}
 G(ip_n,\vecp)$ and $\hat F_{\alpha\alpha'}(ip_n,\vecp)
 = - i\sigma_y\delta{\vs\vs'} F(ip_n,\vecp)$.

Since the quasiparticles are confined to the vicinity of 
the Fermi-surface we expand the normal self-energy around 
the Fermi momentum, $p_F$, to obtain
\bea\label{SPS}
&&\epsilon(p) = \frac{p_F}{m^*}(p-p_F) - \mu^*\,, \\
&&\frac{m^*}{m}  = \left[1+\frac{m}{p_F}
\partial_{p}\Real\Sigma(p)\vert_{p=p_F}\right]^{-1}\, ,
\eea
where $\mu^*\equiv -\epsilon(p_F)+\mu - \Real \Sigma(p_F)$ is the 
effective chemical potential, $m^*$ is the effective mass. 
The dependence of the self-energies on the off-mass shell energy will 
be neglected, i.~e. the wave-function renormalization 
which accounts for the next-to-leading term in the expansion 
around the Fermi-energy is set to unity.
\begin{figure}
\begin{center}
\includegraphics[height=8.0cm,width=9.0cm]{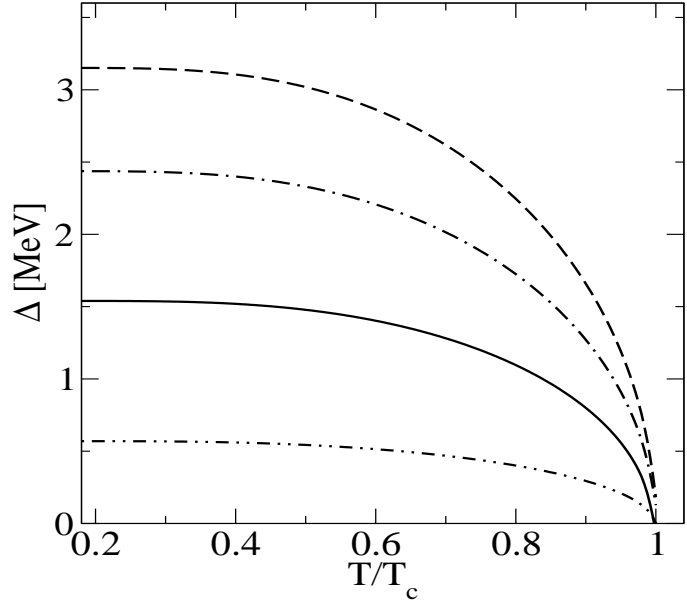}
\end{center}
\caption[]
{
Temperature dependence of the $^1S_0$ pairing gap in neutron matter
for Fermi wave vectors $p_F = 0.4$ ({\it solid line}), 
$0.8$ ({\it dashed line}), 1.2 ({\it dashed-dotted line}) 
and 1.6 ({\it dashed-double-dotted line}).
}
\label{fig1}
\end{figure}
The BCS  mean-field approximation to the anomalous
self-energy can be written in terms of the attractive 
four-point vertex function $\Gamma(p,p')$
in the form
\be\label{GAP1}
\Delta(p) = -2\int\!\frac{d^4p'}{(2\pi)^4} \Gamma(p,p')
~{\rm Im}F(p')f(\omega')\,,
\ee
where $f(\omega) = [1+{\rm exp}(\beta\omega)]^{-1}$ is the Fermi
distribution function and $\beta$ is
the inverse temperature. For the problem of neutron matter 
at subnuclear densities we adopt the standard BCS approach 
and thus approximate the  four-point vertex by the bare interaction
in the $^1S_0$ partial wave channel and the single particle spectrum 
by the on-shell spectrum given by Eq.~(\ref{SPS}).
\begin{table}
\caption{Density dependence of the effective mass, the scalar and spin-spin
interactions, the pairing gap and the critical temperature; the interactions
are given in units of the density of states $\nu(p_F)$.
}
\begin{tabular}{cccccc}
\hline
   $p_F$    & $m^*/m$ &  $v^V$ & $v^A$ & $\Delta(p_F)$ & $T_c$  \\

[fm$^{-1}$] &          &                &               & [MeV]  & [MeV]       \\
\hline 
0.4  & 1.02 & -0.56  & 0.55 & 1.54  & 0.85 \\
0.6  & 1.00 & -0.50  & 0.49 & 2.60  & 1.44\\
0.8  & 0.97 & -0.47  & 0.44 & 3.15  & 1.78\\
1.0  & 0.94 & -0.45  & 0.41 & 3.09  & 1.80\\
1.2  & 0.92 & -0.43  & 0.40 & 2.44  & 1.46\\
1.4  & 0.88 & -0.41  & 0.40 & 1.41  & 0.88\\
1.6  & 0.84 & -0.36  & 0.39 & 0.57  & 0.38\\
\hline
\end{tabular}
\label{tab:1}
\end{table}
With these approximation Eq.~(\ref{GAP1}) reduces to 
\be \label{GAP_ONSHELL}
\Delta(p) = \int\!\frac{d^3 p'}{2(2\pi)^3} 
V(p,p')\frac{\Delta(p')}{\sqrt{\xi(p')^2+\Delta(p')^2}}
\left[1 - 2f(\ep(p'))\right]\, .
\ee
The gap equation is supplemented by the equation for the density 
of the system, 
\bea \label{DENS1}
\rho = -2\int\!\! \frac{d^4 p}{(2\pi)^4} {\rm Im}G(p) f(\omega),
\eea
which determines the chemical potential in a self-consistent manner.
Fig.~\ref{fig1} shows the temperature dependence of the $^1S_0$
pairing gap in neutron matter for several densities parameterized
in terms of the Fermi wave number, $\rho = p_F^3/3\pi^2$. 
The gap at zero temperature and the critical temperature 
for unpairing are listed in the Table~\ref{tab:1}. The dependence
of ratio $\Delta(T)/T$ on $T/T_c$ in this model is non-universal,
i.~e. it depends on the density; this is contrary to the prediction
of the BCS theory with contact pairing interaction.

Within the adopted Fermi-liquid description of neutron matter,
the particle-hole interaction is given by 
\be \label{PH2}
V^{ph}(\vecq) = v^V(\vecq) + v^A(\vecq) (\sigmavec \cdot \sigmavec'),
\ee
where $\sigmavec$ refers to the Pauli matrix. The Landau parameters
$v^V(\vecq)$ and $v^A(\vecq)$ depend on the momentum transfer $\vecq$ in the 
process where both fermion momenta are on the 
Fermi-surface~\cite{NOTE1}.
The tensor and spin-orbit terms are small in neutron matter 
and were neglected in Eq.~(\ref{PH2}).
The full dependence of these parameters on the momentum transfer is 
commonly approximated by a Legendre polynomial with respect to the 
angle formed by the incoming fermions, whereby only the 
leading and next-to-leading order terms contribute significantly.

Table~\ref{tab:1} lists the effective mass, the zeroth order Landau 
parameters in the scalar and spin channels  computed within the 
formalism of Ref.~\cite{DICKHOFF} starting from  the CD Bonn 
potential~\cite{MACHLEIDT}. The solution of 
the gap equation was obtained by applying the iterative method 
with ``running'' cut-off~\cite{SC06} whereby the effective pairing
interaction was approximated by the Gogny DS1 force~\cite{SKMS}.

\section{The pair-breaking processes at one-loop}
\label{sec:3}
The neutrino emissivity (the power of the energy radiated per 
unit volume in neutrino-anti-neutrino pairs) is given 
by~\cite{VS87+MIGDAL90,SEDRAKIAN_DIEPERINK}
\bea\label{EMISSIVITY}
\epsilon_{\nu\anu}&=& - 2\left( \frac{G}{2\sqrt{2}}\right)^2
\int\!\frac{d^3q_1}{(2\pi)^32 \omega_1}
\int\!\frac{d^3 q_2}{(2\pi)^3 2\omega_2}
\int\! dq_0\nonumber\\
&\times& \int\! d^3 \vecq \,\delta (\vecq_1 + \vecq_2 -  \vecq)
\delta(\omega_1+\omega_2-q_{0})\,q_0\nonumber\\
&\times& 
 g(q_0) \Lambda^{\mu\zeta}(q_1,q_2)\Img\,\Pi_{\mu\zeta}(q),
\eea
where $G$ is the weak coupling constant, $q_i = (\omega_i,\vecq_i)$, 
$i=1,2$ are the on-mass-shell four-momenta of neutrinos, 
$g(q_0) = [{\rm exp}(q_0/T) - 1]^{-1}$ 
is the Bose distribution function, $\Pi_{\mu\zeta}(q)$ 
is the retarded polarization tensor, $\Lambda^{\mu\lambda}(q_1,q_2) 
= {\rm Tr}\left[\gamma^{\mu}(1 - \gamma^5)\sla q_1
\gamma^{\nu}(1-\gamma^5)\sla q_2\right]$. Here and below the emissivities
are given per neutrino flavor; the total rate is obtained upon multiplying 
the single flavor rate by the number of neutrino flavors 
within the standard model, $N_f = 3$.
The polarization tensor of baryons at one-loop is shown in
Fig.~\ref{fig:1loop} and  is given analytically by
\be \label{sec2:eq1}
\Pi^{V/A}(q) = T \sum_{\sigma, p}
\left[ G(p)G(p+q) \mp F(p)\dF (p+q) \right],
\ee
where $p = (ip_0,\vecp)$. The upper/lower signs correspond to vector 
current  ($V$) and axial vector  current ($A$) couplings.
In writing Eq.~(\ref{sec2:eq1}) we assumed that baryons carry the same 
isospin quantum number.
\begin{widetext}
\begin{figure}[tb]
\begin{center}
\includegraphics[height=2.cm,width=14cm]{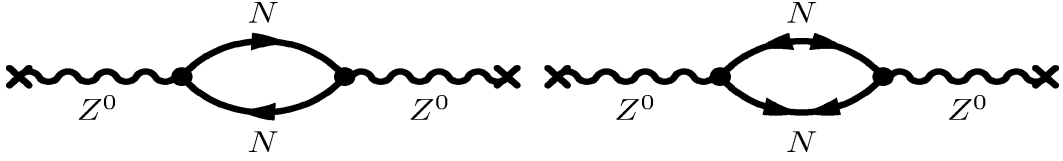}
\end{center}
\begin{minipage}{16cm}
\caption{ The one-loop contribution to the polarization tensor
in the superfluid matter; solid lines refer to the baryon propagators, 
wavy lines to the (amputated) $Z^0$ propagator.
}\label{fig:1loop}
\end{minipage}
\end{figure} 
\noindent
Performing the Matsubara sums in Eq. (\ref{sec2:eq1})
(see the Appendix) we obtain
\bea \label{sec2:eq2}
\Pi^{V/A}(q) &=& \sum_{\sigma\vecp} 
\left[f(\ep_p) -f(\ep_k)\right] \left(
\frac{A_{\mp}}{iq+\ep_p-\ep_k} -\frac{B_{\mp}}
{iq-\ep_p+\ep_k}\right)\nonumber\\
&+&\sum_{\sigma\vecp} \left[f(-\ep_p) -f(\ep_k)\right] \left(
\frac{C_{\mp}}{iq-\ep_k-\ep_p}-\frac{D_{\mp}}{iq+\ep_p+\ep_k}
\right),
\eea
where $k = p +q$, $A_{\mp} = u_p^2u_k^2 \mp h$, 
      $B_{\mp} = v_p^2v_k^2 \mp h$,
      $C_{\mp} = u_k^2v_p^2 \pm h$,
      $D_{\mp} = u_p^2v_k^2 \pm h$, 
$h = u_pu_kv_pv_k$.
The emissivity (\ref{EMISSIVITY}) requires 
the  imaginary part of the polarization  tensor which after analytical 
continuation in Eq.~(\ref{sec2:eq2}) becomes [note that $f(-\ep_p) =
1-f(\ep_p)]$
\bea\label{sec2:eq3}
\Img\,\Pi^{V/A}(q) &=& -\pi \sum_{\sigma\vecp} 
\left[f(\ep_p)-f(\ep_k)\right]
 (A_{\mp}+B_{\mp})~\delta(\omega+\ep_p-\ep_k)
       \nonumber \\
&-&\pi \sum_{\sigma\vecp}
\left[f(-\ep_p)-f(\ep_k)\right]
\left[ C_{\mp}~\delta(\omega-\ep_p-\ep_k)
       -D_{\mp}~\delta(\omega+\ep_p+\ep_k)
\right].
\eea
To obtain the first line we used the fact that 
the quasiparticle spectra are invariant under spatial 
reflections, i.~e., $\ep(-\vecp) = \ep(\vecp)$.
\end{widetext}
The first line in Eq.~(\ref{sec2:eq3}) 
corresponds to the process of scattering
where a quasiparticle is promoted out of the condensate into 
an excited state, or inversely, an excitation merges with 
the condensate.  The corresponding piece of the response 
function $\Img\Pi^{V/A}(q)$ vanishes for small momentum transfers. 
Indeed neutrino energies are of the order of temperature
which implies that their wave vectors
$q [{\rm fm}^{-1}]
\sim\omega_{\nu}/\hbar c\sim T/\hbar c \sim 1/197.3 \ll 1$. 
On the other hand the neutron wave vectors   
$p \sim p_F \sim 1$ fm$^{-1}$. Upon expanding the argument 
of the delta-function with respect to $\vert q\vert$ 
around the point $q = 0 $ one finds
\begin{equation}
\label{eq:3}
\omega
- q\,\frac{\xi_p}{\ep_p}
\frac{\partial \xi_{p+q}}{\partial q}\Big\vert_{q=0} - 
q\,\frac{\Delta_p}{\ep_p}
\frac{\partial\Delta_{p+q}}{\partial q}\Big\vert_{q=0}= 0.
\end{equation}
If we assume that $\Delta \neq \Delta(p)$ the third term on 
the l.~h. side  vanishes. It  follows that 
$x\equiv (\vecp\cdot \vecq)/pq = (\epsilon_p/\xi_p)(\omega/v q)\le 1$, 
where $v~ (\sim v_F)$ is the baryon (Fermi) velocity.
For on mass-shell neutrinos $\omega = cq$ and the latter condition 
can not be satisfied, i.~e. the scattering contribution can 
be neglected. The non-locality in the momentum or frequency domains
of the gap function will alter this conclusion, but will require
a specific model of the momentum and frequency dependence of the gap 
function (for $S$-wave interactions the momentum and energy
dependences are described respectively 
in Refs.~\cite{LOMB3,SKMS,KUCKEI,MUDICK} 
and \cite{LOMB1,SEDRAKIAN03,LOMB2}). 
The second line in Eq. (\ref{sec2:eq3}) describes the process 
of pair-breaking and recombination, i.~e., excitation of pairs 
of quasiparticles out of the condensate, and inversely, 
restoration of a pair within the condensate. Since we are interested
in the emission process we shall keep only the terms that 
do not vanish for $\omega > 0$; then, the pair-braking contribution 
is given by 

\bea\label{sec2:eq4}
\Img\,\Pi^{V/A}(q) = -\pi \sum_{\sigma\vecp}
\left[f(-\ep_p)-f(\ep_k)\right]C_{\mp}.
\eea
In the limit $\vecq\to 0$ and assuming that $\Delta\neq \Delta(p)$ 
the integrations in  Eq.~(\ref{sec2:eq4}) can be performed analytically.
One finds
\bea\label{sec2:eq5}
\Img\,\Pi^{V}(q) &=& - 2\pi \nu(p_F)
g(\omega)^{-1} f\left(\frac{\omega}{2}\right)^2
\left(\frac{\Delta^2}{\omega^2}\right)
\frac{\omega}{\sqrt{\omega^2-4\Delta^2}}\theta(\omega-2\Delta),\\
\Img\,\Pi^{A}(q) &\simeq & 0 + O\left(\frac{v_F^2}{c^2}\right),
\eea
where  $\nu(p_F) = m^* p_F/2\pi^2$ is the density of states ($\hbar = 1$)
and $\theta$ is the Heaviside step function; the explicit 
form of the $O\left({v_F^2}/{c^2}\right)$ contribution to the 
axial current response is given in Ref.~\cite{FRS78}. 
Note the threshold behavior of the 
vector current response, which is finite for frequencies that 
are large compared to $2\Delta$ - the energy needed to break a pair.

Upon substituting Eq.~(\ref{sec2:eq5}) in Eq.~(\ref{EMISSIVITY}) 
and carrying out the phase-space integrals we obtain the
emissivity per neutrino flavor~\cite{FRS78,VS87+MIGDAL90}
\be 
\epsilon^{\rm 1-loop}_{\nu\bar\nu} 
= \epsilon_{0}~  z^2~\int_{2z}^{\infty}\!\! dx 
\frac{x^5}{\sqrt{x^2-4z^2}}~f\left(\frac{x}{2}\right)^2
\ee
where $z =\Delta(T)/T$ and 
\be \label{EMISSIVITY2}
\epsilon_{0} =  \frac{\nu(p_F)G^2 c_V^2}{60~\pi^3}T^7.
\ee 

\section{Weak interaction vertices }
\label{sec:4}
\begin{figure}[t]
\begin{center}
\includegraphics[height=4.5cm,width=15cm]{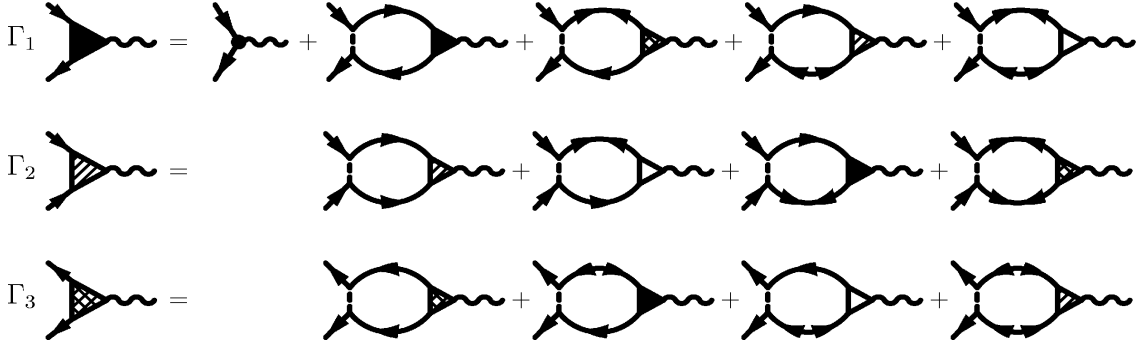}
\end{center}
\caption[]
{
Coupled integral equations for the effective weak vertices
in superfluid baryonic matter. The ``normal'' $\Gamma_1$ vertex 
(full triangle) and two ``anomalous'' vertices $\Gamma_2$ 
(hatched) and $\Gamma_3$ (shaded triangle) are shown 
explicitly, the fourth vertex (empty triangle) is obtained by 
interchanging the particle and hole lines in the first line.
The anomalous vertices vanish in the normal state.
}
\label{fig2}
\end{figure}
Consider the weak vector and axial vector vertices in the
nuclear medium featuring a condensate. Because the particle-hole
interactions in the medium (which are represented by the Landau
parameters in Table~\ref{tab:1}) are not small, the
vertex renormalization requires a summation of an infinite 
number of particle-hole loops. There are four topologically 
non-equivalent vertices in the superfluid state in general, 
but, as we shall see, the particle-hole symmetry reduces their 
number by one. Since neutrons pair in an isospin-1 
state (neutron-proton pairing is 
unimportant for large asymmetries~\cite{ASYM}) 
we shall suppress the isospin indices.  The expansion of the Landau 
parameters in Legendre polynomials will be truncated at the leading 
order (the next-to-leading order terms are suppressed
by powers of $v_F/c$). Thus, the effective particle-hole interaction 
is given by
\be \label{LP}
v = v^V + v^A (\sigmavec \cdot \sigmavec').
\ee
The integral equation defining the effective weak vertices, which we
write in an operator form,  are given by 
\bea\label{g1}
\hat\Gamma_1^{a} &=& \Gamma_0^{a} 
             +  v^{a} (G\Gamma_1^{a}G
             +       \hat F\Gamma_3^{a}G
             +       G\Gamma_2^{a}\hat F
             +       \hat F\Gamma_4^{a}\hat F),  \\ 
\label{g2}
\hat\Gamma_2^{a} &=&  \quad \quad v^{a} (G\Gamma_2^{a}G^{\dagger}
             +       \hat F\Gamma_4^{a}G^{\dagger}
             +       G\Gamma_1^{a}\hat F
             +       \hat F\Gamma_3^{a}\hat F), \\ 
\label{g3}
\hat\Gamma_3^{a} &=&  \quad \quad v^{a} (G^{\dagger}\Gamma_3^{a}G
             +       \hat F\Gamma_1^{a}G
             +       G^{\dagger}\Gamma_4^{a}\hat F
             +       \hat F\Gamma_2^{a}\hat F),
\eea
and are displayed diagrammatically in Fig.~\ref{fig2}. Here $\hat F = 
-i\sigma_y F$,  $v^{a}$  with $a \in V, A$ are defined in 
Eq.~(\ref{LP}), while $\Gamma_0^{V} = 1$ and  
$\Gamma_0^{A} = \sigmavec $. The fourth 
integral equation for the vertex 
$\Gamma_4^{a}$ follows upon interchanging in Eq.~(\ref{g1})
particle and hole propagators $G \leftrightarrow G^{\dagger}$. 
The momentum space representation of Eq.~(\ref{g1}) reads
\bea
\hat \Gamma_1^{a}(q) = \Gamma_0^{a} 
             &+&  v^{a} \int \frac{d^4p}{(2\pi)^4}
\Bigr[G(p)\Gamma_1^{a}(q)G(p+q)
             +       G(p)\Gamma_2^{a}(q)\hat F(p+q)\nonumber\\
&+&       F\hat (p)\Gamma_3^{a}(q)G(p+q)
             +       \hat F(p)\Gamma_4^{a}(q)\hat F(p+q)\Bigl],
\eea
with similar expressions for Eqs. (\ref{g2}) and (\ref{g3}).
Even though the driving interactions are local in 
time, the resummations at finite temperatures lead to 
time-retarded interactions, which imply that 
the effective vertices in Eqs.~(\ref{g1})-(\ref{g3}) 
are complex in general. Considering the scalar 
interaction $\Gamma_0^V$ we obtain 
\be\label{G1}
v^V\left(\begin{array}{ccc}
(v^V)^{-1}- [\Pi_{GG}(q)- \Pi_{FF}(q)] &\Pi_{GF}(q) & \Pi_{FG}(q)\\
-2\Pi_{GF}(q) & (v^V)^{-1}-\Pi_{GG\dagger }(q) &\Pi_{FF}(q)\\
-2\Pi_{FG}(q) & \Pi_{FF}(q) & (v^V)^{-1}-\Pi_{G\dagger G}(q)
\end{array}\right)\left(\begin{array}{c}
\Gamma^V_1(q)\\
\Gamma^V_2(q)\\
\Gamma^V_3(q) 
\end{array}\right) = \left(\begin{array}{c}
\Gamma^V_0\\
0\\
0 
\end{array}\right),
\ee
where, using the abbreviations $X,X'\in G,F,G^{\dagger}$, 
we defined (see the Appendix)
\be \label{PIDEF}
\Pi_{XX'}(q) = \int\frac{d^3\vecp}{(2\pi)^3}L_{XX'}(q,\vecp)
= \int\frac{d^3\vecp}{(2\pi)^3}\sum_{ip} X(p)X'(p+q).
\ee
The loop integrals 
posses the following symmetries: $L_{GG} = L_{G^{\dagger}G^{\dagger}}$, 
$L_{GF} = L_{FG^{\dagger}}$ and $L_{FG} = L_{G^{\dagger}F}$, which
imply that $\Gamma_1^{V} = \Gamma_4^{V}$ and $\Gamma_1^{A} = -\Gamma_4^{A}$.
For energy-momentum independent interactions the integral equation 
(\ref{G1}) reduces to three coupled linear equation for 
the complex functions $\Gamma^V_i$, $i = 1,2,3$. The details of
the computation of the coefficients in Eqs.~(\ref{G1})
are relegated to the Appendix. 

In the case of axial-vector interactions the effective vertices 
are given by
\be\label{G2}
v^A\left(\begin{array}{ccc}
(v^A)^{-1}& \Pi_{GF}(q) & \Pi_{FG}(q)\\
0 &(v^A)^{-1}-\Pi_{GG\dagger }(q) &\Pi_{FF}(q)\\
0 & \Pi_{FF}(q) & (v^A)^{-1}-\Pi_{G\dagger G}(q)
\end{array}\right)\left(\begin{array}{c}
\Gamma^A_1(q)\\
\Gamma^A_2(q)\\
\Gamma^A_3(q) 
\end{array}\right) = \left(\begin{array}{c}
\Gamma^A_0\\
0\\
0 
\end{array}\right).
\ee
This equation has a ``trivial'' solution $\Gamma^A_1 = 1$ 
and $\Gamma^A_2 = \Gamma^A_3 = 0$ to the leading order in the 
$v_F^2/c^2$ expansion and we shall not consider it further.
In the weak-coupling BCS limit there exist
{\it approximate} constraints among the loop integrals 
\be \label{AP_DEG}
L_{GF} = -L_{FG},\quad\quad L_{G^{\dagger}G} =  L_{GG^{\dagger}},
\ee
which allow us to reduce the number of equations in the set (\ref{G1})
from three to two. The relations (\ref{AP_DEG}) are exact at the
threshold $\omega = 2\Delta$ and hold approximately for systems 
with strong degeneracy (corrections being suppressed by powers of
the ratio of the temperature over the Fermi-energy). It follows then 
from Eq.~(\ref{G1}) that $\Gamma_2^V = -\Gamma_3^V$. The solutions for 
the remaining vertices $\Gamma_1$ and $\Gamma_2$
are convenient to express through linear combinations of the 
polarization tensors (\ref{PIDEF}), defined as 
\bea \label{A}
{\cal A}(q) &=& \Pi_{GG}(q)-\Pi_{FF}(q)
=\int\!\!\frac{d^3p}{(2\pi)^3}
{\cal F}_p^{-}(\vecq)
{\cal L}_p(\omega,\vecq),\\
\label{B}
{\cal B}(q) &=& 2\Pi_{FG}(q) = - \omega\Delta\int\!\!\frac{d^3p}{(2\pi)^3}
\frac{{\cal L}_p(\omega,\vecq)}{\ep_p},\\
\label{C}
{\cal C}(q) &=& \Pi_{GG^{\dagger}}(q)+\Pi_{FF}(q)-(v^V)^{-1}
= \int\!\!\frac{d^3p}{(2\pi)^3}\left[
2\ep_p {\cal L}_p(0, 0) - 
{\cal F}_p^{+}(\vecq){\cal L}_p(\omega,\vecq)\right],
\eea
where ($\veck = \vecp+\vecq$)
\bea 
{\cal F}_p^{\pm}(\vecq) &=& \left(\frac{\ep_p+\ep_k}{2}\right)
\left(1\pm\frac{\xi_p\xi_k}{\ep_p\ep_k}
+\frac{\Delta^2}{\ep_p\ep_k}\right), \\
{\cal L}_p(\omega,\vecq) &=&
\frac{1-f(\ep_p)-f(\ep_k)}{\omega^2-(\ep_p+\ep_k)^2+i\delta} .
\eea
In obtaining Eq.~(\ref{C}) we used the fact that 
\be \label{GAP_CUT}
1+v^V\int^{\Lambda} \frac{d^4p}{(2\pi)^4}
(\Pi_{GG^{\dagger}}+\Pi_{FF})\Bigr|_{\omega =
  0, \vecq = 0} = 0,
\ee
where $\Lambda$ is a three-dimensional ultraviolet cut-off on the momentum 
integration, which is required for regularization of the gap 
equation (\ref{GAP_CUT}). $\Lambda$ may be adjusted to 
reproduce the gaps obtained from  finite range interactions 
in Sec.~\ref{sec:2}, but such an adjustment is not required as 
long as the constraint (\ref{GAP_CUT}) is taken into account
in finding the solutions for the vertex functions.

The vertex functions expressed through the combination
(\ref{A})-(\ref{C}) read
\bea 
\Gamma_1 (q) &=& \frac{{\cal C}(q)}{{\cal C}(q)
-v^V[{\cal A}(q){\cal C}(q)
+{\cal B}(q)^2]},\\
\Gamma_2 (q) &=& -\frac{{\cal B}(q)}{{\cal C}(q)
-v^V[{\cal A}(q){\cal C}(q)
+{\cal B}(q)^2]}.
\eea
The poles of the functions $\Gamma_1 (q)$ and $\Gamma_2 (q)$ 
determine the collective excitations of the $S$ wave superfluid, 
which we will discuss in the next section.


\section{The Full polarization tensor and the emissivity}
\label{sec:5}
\begin{figure}[t]
\begin{center}
\includegraphics[height=2.cm,width=\linewidth]{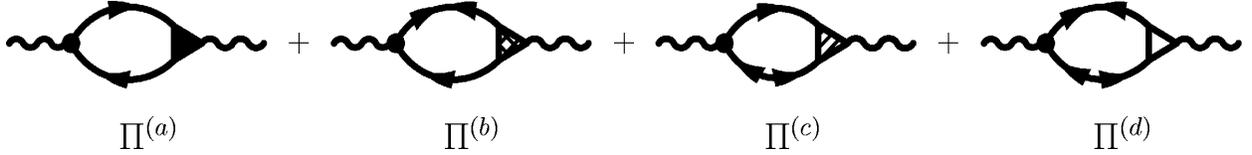}
\end{center}
\caption[]
{The sum of polarization tensors that 
 contribute to the neutrino emission rate.  The contributions
form $\Pi^{(b)}(q)$ and $\Pi^{(c)}(q)$ vanish at one-loop.
}
\label{fig3}
\end{figure}
Having determined the effective vertices 
$\Gamma_{1}^V(\omega)$ and $\Gamma_{2}^V(\omega)$, 
we now construct the complete polarization 
tensor, which is given by the sum of diagrams in Fig.~\ref{fig3}. 
Summing the contributions we obtain the full polarization tensor
expressed through the combinations (\ref{A})-(\ref{C}):
\be \label{FULL_PI}
\Pi^V (q) = \frac{{\cal A}(q){\cal C}(q)+{\cal B}(q)^2}
{{\cal C}(q) -v^V[{\cal A}(q){\cal C}(q)+{\cal B}(q)^2]}.
\ee
Eq. (\ref{FULL_PI}) is our central result valid for arbitrary momentum 
transfers. It can be used to compute the emissivity directly, but 
it is illuminating to work with its small $q$ expansion.
The leading order term in this expansion vanishes, i.~e.  
the full polarization tensor vanishes when 
$\vecq =  0$ (note that this is contrary to the one-loop 
polarization tensor, which is finite in this limit,
see Eq. (\ref{sec2:eq2}) and the 
following discussion.) Indeed, taking the limit $\vecq \to 0$ in 
Eqs.~(\ref{A})-(\ref{C}) and substituting in Eq.~(\ref{FULL_PI}) 
we find that
\be \label{LIM}
\lim_{\vecq\to 0} \Pi^V (\vecq,\omega) = 0.
\ee
This result is consistent with the $f$-sum rule~\cite{NEGELE_ORLAND}
\be 
\lim_{\vecq\to 0}\int  d\omega ~\omega ~{\rm Im}\Pi^V(\vecq,\omega) = 0,
\ee
which follows directly from (\ref{LIM}). The opposite need not to be 
true for arbitrary functional dependence of ${\rm Im}\Pi^V(\vecq,\omega)$
on frequency, but is the case in practice. The reason is that 
the causal polarization tensors are odd functions of frequency.
Furthermore, since they should correspond to stable collective modes
the condition $\omega~{\rm Im}\Pi^V(\vecq,\omega) \ge 0$  is 
satisfied, which combined with the $f$-sum rule leads 
back to Eq.~(\ref{LIM}).

Consider now the next-to-leading order terms. Keeping only the
leading order ${\cal F}^{\pm}_p(0)$ function in Eqs. (\ref{A})-(\ref{C}) 
one finds that the polarization function vanishes order by order 
in the expansion of the function ${\cal L}_p(\omega,\vecq)$; thus instead 
of straightforward expansions of kernels in Eqs.~(\ref{A})-(\ref{C}), 
we shall expand only the functions ${\cal F}^{\pm}_p(\vecq)$:
\bea 
{\cal F}_p^{+}(\vecq) 
= \left(\frac{\ep_p+\ep_k}{2}\right)
\left(1+\frac{\xi_p\xi_k}{\ep_p\ep_k}
+\frac{\Delta^2}{\ep_p\ep_k}\right)  
=  2\ep_p +\frac{\xi_p\xi_q}{\ep_p}
\eea
and 
\bea 
{\cal F}_p^{-}(\vecq) 
= \left(\frac{\ep_p+\ep_k}{2}\right)
\left(1-\frac{\xi_p\xi_k}{\ep_p\ep_k}
+\frac{\Delta^2}{\ep_p\ep_k}\right)  
=  \frac{2\Delta^2}{\ep_p} - \frac{\Delta^2\xi_p\xi_q}{\ep_p^3},
\eea
where $\xi_q = q^2/2m$ is the nucleon recoil (the linear in $q$ terms 
are omitted since they vanish upon angle integrations). Substituting these 
expressions back into Eqs.~(\ref{A})-(\ref{C}) one finds
\bea\label{Abis} 
{\cal A}(q) &=& 2\Delta^2~I_0(q)-\Delta^2\xi_q ~I_A(q),\\
\label{Bbis} 
{\cal B}(q) &=& - \omega\Delta~I_0(q),\\
\label{Cbis} 
{\cal C}(q) &=& - \frac{\omega^2}{2}~I_0(q)
+\xi_q~I_C(q),
\eea
where we defined the following integrals
\bea 
I_0(q) =\int\!\!\frac{d^3p}{(2\pi)^3}
\frac{1}{2\ep_p}{\cal L}_p(\omega,\vecq), \\
I_A(q) =\int\!\!\frac{d^3p}{(2\pi)^3}\frac{\xi_p}
{\ep_p^3}{\cal L}_p(\omega,\vecq), \\
I_C(q) =\int\!\!\frac{d^3p}{(2\pi)^3}\frac{\xi_p}
{\ep_p}{\cal L}_p(\omega,\vecq) .
\eea
The pole(s) of the polarization tensor (or equivalently of the 
vertex function) determine the dispersion relation of 
the collective excitations:
\bea 
{\cal D}(q) &=& {\cal C}(q) -v^V[{\cal A}(q){\cal C}(q)+{\cal B}(q)^2]
\nonumber\\
 &\simeq&  {\cal C}_0(q)+{\cal C}_1(q) - 
v^V[{\cal A}_1(q){\cal C}_0(q)+{\cal A}_0(q){\cal C}_1(q)] =0,
\eea
where the subscripts 0 and 1 refer to the leading and next to leading
order terms in the small $q$ expansion. Substituting here the expressions
(\ref{Abis})-(\ref{Cbis}) we obtain the dispersion relation for 
the acoustic modes 
\be 
\omega^2 =  c^2q^2,
\ee
where the sound velocity is defined as 
\bea 
c^2 &=& \frac{I_C}{2m^*}\left\{ I_0^{-1} - 
4v^V \Delta^2\left[1+ \omega^2~\frac{I_A}{4I_C}\right]\right\}.
\eea
When $v^V = 0$, the sound velocity at zero temperature  
should reduce to that of non-interacting 
Fermi-gas, $c^2 = v_F^2/3$. It is seen that the ratio 
$I_C/I_0\sim\xi_p$; since the average of $ \xi_p\sim p_Fv_F$
one recovers the scaling $c^2\sim v_F^2$.

\begin{figure}[t]
\begin{center}
\includegraphics[height=10.cm,width=12.cm,angle=0]{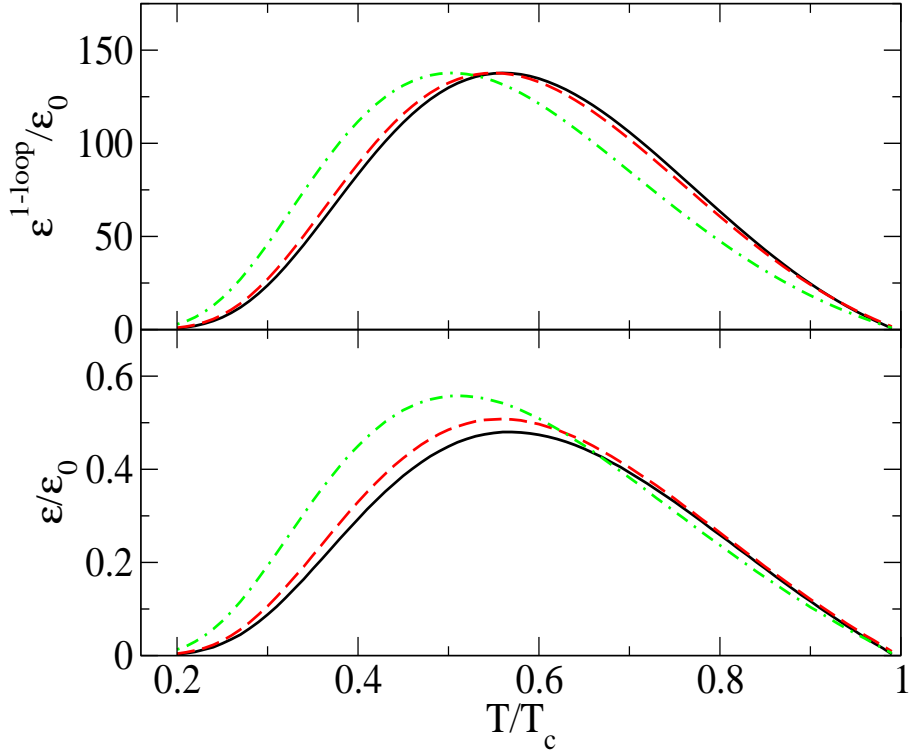}
\end{center}
\caption[]
{(Color online) Dependence of neutrino emissivity in units of $\epsilon_0$
for one-loop  ({\it upper panel}) and full ({\it lower panel})
polarization tensor on the reduced temperature 
for Fermi wave vectors $p_F[{\rm fm}^{-1}] 
= 0.4$ (solid),~ 0.8~(dashed, red online),
and 1.4 (dashed-dotted, green online). 
}
\label{fig4}
\end{figure}

At the next-to-leading order in small-$q$ 
expansion the polarization tensor is given by 
\bea 
\Pi_1^V (q) = {\cal A}_1(q) +
\frac{{\cal A}_0(q)}{{\cal C}_0(q)}~{\cal C}_1(q)
=-\Delta^2\xi_q\left[ I_A(q)  +\frac{4}{\omega^2}I_C(q)\right].
\eea 
It is sufficient to evaluate the integrals in the limit $q =0$ at 
the order we are working; then the imaginary part of the 
polarization tensor can be evaluated analytically
\be
{\rm Im}\Pi_{1}(\omega) =  
-8\pi\nu(p_F)\frac{\Delta^2\xi_q }{\omega^3}
g(\omega)^{-1}f\left(\frac{\omega}{2}\right)^2\theta(\omega -2\Delta).
\ee
The neutrino emissivity to the leading $O(q^2)$ order is 
\be \label{EMISSIVITY3}
\epsilon_{\nu\bar\nu} = \epsilon_{0}~
z^2 \frac{T}{m^*}\int_{2z}^{\infty} dx~ x^5   f\left(\frac{x}{2}\right)^2.
\ee
Compared to the one-loop result, the emissivity is suppressed by 
a factor $T/m^*$ and shows a different functional dependence on the 
frequency; this difference arises from the  parametric suppression
of the full polarization tensor $\Pi_1^V (q)\sim \xi_q$ and the 
fact that in the bremsstrahlung process $q\sim T$. The  emissivity
is thus suppressed compared to the one-loop result, roughly 
by a factor $4\times 10^{-3}$, as illustrated in  Fig.~\ref{fig4}.
Note that the density dependence in  Fig.~\ref{fig4} 
arises from the density dependence of 
the ratio $\Delta(T)/T$ as a function of $T/T_c$ (in the BCS theory
with contact interaction the ratio $\epsilon_{\nu\bar\nu}/\epsilon_{0}$
is universal).
For $T\to T_c$ the both rates vanish, consistent with the observation 
that the pair bremsstrahlung is absent in normal matter for on-shell 
baryons. At small $T\le 0.3 T_c$ the rates are suppressed exponentially 
as ${\rm exp}(-\Delta/T)$. At intermediate temperatures 
the emissivities that include vertex corrections should be
scaled down from their one-loop counterparts 
roughly by the  factor $\sim 4\times 10^{-3}$  quoted above.

\section{Summary}
\label{sec:6}

The neutrino losses via neutrino pair bremsstrahlung 
from neutron superfluid at subnuclear densities have been studied
within a microscopic model based on the Fermi-liquid and BCS
theories. In our model the modifications of the weak interaction rates 
in the nuclear medium are taken into account by summing infinite 
series of irreduceable particle-hole diagrams in 
terms of a contact (momentum and energy independent) driving 
interaction. These modification are
embodied in the vertex functions that are computed from 
polarization tensor components corresponding to the pair-breaking 
process within the kinematical domain $\omega\in [2\Delta;\infty]$ 
and small-$q$ expansion. Special attention was paid to preserving the 
dispersion relations for the polarization tensor components, 
that are implied by the unitarity of the $S$-matrix 
for the emission process. The leading order contribution to the 
rate arises from $O(q^2$) contribution to the polarization 
tensor, which vanishes identically when $\vecq = 0$.
We find neutrino loss rates that are suppressed compared to 
the one-loop results by a factor of the order $4\times 10^{-3}$. 
The magnitude of the suppression differs from the one
predicted in Ref.~\cite{LP} due to two factors: (i)~the vertex functions 
used herein include the finite temperature effects, which guarantees their
consistency with the associated polarization 
tensors; (ii)~the full polarization tensor is derived in a 
different expansion (small-$q$ rather than the $v_F/c$ expansion).

The modifications to the neutrino emission rate through pair-breaking 
process found above call for a detail reassessment of their role in 
the late-time cooling of neutron stars. Even though the rates are 
suppressed substantially, their strong dependence on the value of the 
pairing gap does not rule out the possibility that for large enough 
gaps the rates will become comparable to those of the competing 
processes. 

While we concentrated above on the neutral current interactions, 
our formalism can be adapted to compute the rates of charge-current 
Urca process $n \to p + e + \nu_e$ beyond  the one-loop 
rate~\cite{SEDRAKIAN05}. These should be subject to strong correction 
due to the tensor neutron-proton correlations. Vertex corrections 
could be important in the related problem of neutrino emission and 
propagation in quark matter~\cite{KUNDU_REDDY}, 
where one-loop results for a number of processes became available 
recently~~\cite{JRS,Wang:2006tg,
Schmitt:2005wg,Alford:2006gy,SAD,ANGLANI}.

\section*{Acknowledgments}
We are grateful to Sanjay Reddy for helpful communications and 
for drawing our attention to the sum rules. This work was, in part,
supported by the Deutsche Forschungsgemainschaft through
the SFB 382.

\appendix

\section{Loop integrals}
Here we quote the results for the loop integrals and polarization 
tensors that have been used in the main text. The loop integrals 
are defined as convolution products of the Matsubara Green's 
functions (here $\veck = \vecp+\vecq$) 
\bea 
\label{A1}
L_{GG}(q,\vecp) &=& T\sum_{ip} G(ip,\vecp) G(ip+iq, \veck)\nonumber\\
&=&  \left\{ \frac{u_p^2u_k^2}{iq+\ep_p-\ep_k}- 
             \frac{v_p^2v_k^2}{iq-\ep_p+\ep_k}
     \right\}\left[f(\ep_p)-f(\ep_k)\right]\nonumber\\
&+&  \left\{ \frac{u_k^2v_p^2}{iq-\ep_p-\ep_k}- 
             \frac{u_p^2v_k^2}{iq+\ep_p+\ep_k}
     \right\}\left[f(-\ep_p)-f(\ep_k)\right],
\eea
\bea
\label{A2}
L_{FG}(q,\vecp) &=& T\sum_{ip} F(ip,\vecp) G (ip+iq, \veck)\nonumber\\
&=& -u_pv_p \left[ \frac{u_k^2}{iq+\ep_p-\ep_k}+ 
             \frac{v_k^2}{iq-\ep_p+\ep_k}
     \right]\left[f(\ep_p)-f(\ep_k)\right]\nonumber\\
&+& u_pv_p \left[ \frac{u_k^2}{iq-\ep_p-\ep_k}+
             \frac{v_k^2}{iq+\ep_p+\ep_k}
     \right]\left[f(-\ep_p)-f(\ep_k)\right],
\eea
\bea
\label{A4}
L_{FF}(q,\vecp) &=& T\sum_{ip} F(ip,\vecp) \dF (ip+iq, \veck)\nonumber\\
&=& u_pu_kv_pv_k \Biggl\{\left[\frac{1}{iq+\ep_p-\ep_k}- 
             \frac{1}{iq-\ep_p+\ep_k}
     \right]\left[f(\ep_p)-f(\ep_k)\right]\nonumber\\
&+&  \left[ \frac{1}{iq+\ep_p+\ep_k}- 
             \frac{1}{iq-\ep_p-\ep_k}
     \right]\left[f(-\ep_p)-f(\ep_k) \right]\Biggr\},
\eea
\bea
\label{A5}
L_{G^{\dagger}G}(q,\vecp) &=& T\sum_{ip} G^{\dagger}(ip,\vecp) G (ip+iq, \veck)\nonumber\\
&=& - \left[ \frac{u_k^2v_p^2}{iq+\ep_p-\ep_k} -
             \frac{u_p^2v_k^2}{iq-\ep_p+\ep_k}
     \right]\left[f(\ep_p)-f(\ep_k)\right]\nonumber\\
&-&  \left[ \frac{u_p^2u_k^2}{iq-\ep_p-\ep_k}-
             \frac{v_p^2v_k^2}{iq+\ep_p+\ep_k}
     \right]\left[f(-\ep_p)-f(\ep_k)\right],
\eea
\bea
\label{A7}
L_{FG^{\dagger}}(q,\vecp) &=& T\sum_{ip} F(ip,\vecp)
 G^{\dagger} (ip+iq, \veck)\nonumber\\
&=& u_pv_p \left[\frac{v_k^2}{iq+\ep_p-\ep_k} +
             \frac{u_k^2}{iq-\ep_p+\ep_k}
     \right]\left[f(\ep_p)-f(\ep_k)\right]\nonumber\\
&-&u_pv_p  \left[ \frac{v_k^2}{iq-\ep_p-\ep_k}+
             \frac{u_k^2}{iq+\ep_p+\ep_k}
     \right]\left[f(-\ep_p)-f(\ep_k)\right].
\eea
The remainder loop integrals are obtained from those defined 
above through the relations
\bea\label{PARITY_L1}
L_{G^{\dagger}G^{\dagger}}(iq,\vecp) = L_{GG}(-iq,\vecp),\quad
L_{GF}(iq,\vecp) = L_{FG}(-iq,\vecp),\\
\label{PARITY_L2}
L_{GG^{\dagger}}(iq,\vecp) = L_{G^{\dagger}G}(-iq,\vecp),\quad 
L_{G^{\dagger} F}(iq,\vecp) = L_{FG^{\dagger}}(-iq,\vecp),
\eea
where, except for the first relation, we used the 
fact that the quasiparticle spectrum is reflection 
invariant, $\ep(-\vecp) = \ep(\vecp)$. This property 
implies that the arguments of the functions can be interchanged
$\veck\leftrightarrow \vecp$ without changing the result. 
Furthermore, upon performing the substitution $\vecp \to -\vecp -\vecq$
one finds that 
\be 
L_{GG}(iq,\vecp) = L_{G^{\dagger}G^{\dagger}}(iq,\vecp) ,\quad
L_{GF}(iq,\vecp) = L_{FG^{\dagger}}(iq,\vecp),\quad
L_{FG}(iq,\vecp) = L_{G^{\dagger}F}(iq,\vecp).
\ee
The retarded polarization tensor is obtained by analytical 
continuation in the loop integrals 
$L_{XX'}(iq,\vecp) = L_{XX'}(\omega+i\delta,\vecp)$ 
and by subsequent integration over the three-momentum $\vecp$ 
according to the Eq.~(\ref{PIDEF}).


\begin{thebibliography}{99}
\bibitem{NOTE1} In the common notation these quantities correspond 
               to the $f$ and $g$ Landau parameters; the present
               notation anticipates that these constants are 
               the driving terms in the vector and axial-vector channels.
\end{thebibliography}
\end{document}